# The energy sensor AMPK: adaptations to exercise, nutritional and hormonal signals

Benoit VIOLLET[1,2,3]

[1]INSERM, U1016, Institut Cochin, Paris, France
[2]CNRS, UMR8104, Paris, France
[3]Université Paris Descartes, Sorbonne Paris Cité, France

Correspondence:
Benoit Viollet
Institut Cochin, Inserm U1016, Département Endocrinologie, Métabolisme et Diabète, 24 rue du faubourg Saint-Jacques 7504 Paris, France.
benoit.viollet@inserm.fr
Phone: + 33 1 44 41 24 01; Fax: +33 1 44 41 24 21

**Abstract**

To sustain metabolism, intracellular ATP concentration must be regulated within an appropriate range. This coordination is achieved through the function of the AMP-activated protein kinase (AMPK), a cellular "fuel gauge" that is expressed in essentially all eukaryotic cells as heterotrimeric complexes containing catalytic α subunits and regulatory β and γ subunits. When cellular energy status has been compromised, AMPK is activated by increases in AMP:ATP or ADP:ATP ratios and acts to restore energy homeostasis by stimulating energy production via catabolic pathways while decreasing non-essential energy-consuming pathways. Although the primary function of AMPK is to regulate energy homeostasis at a cell-autonomous level, in multicellular organisms, the AMPK system has evolved to interact with hormones to regulate energy intake and expenditure at the whole body level. Thus, AMPK functions as a signaling hub, coordinating anabolic and catabolic pathways to balance nutrient supply with energy demand at both the cellular and whole-body levels. AMPK is activated by various metabolic stresses such as ischemia or hypoxia or glucose deprivation and has both acute and long-term effects on metabolic pathways and key cellular functions. In addition, AMPK appears to be a major sensor of energy demand in exercising muscle and acts both as a multitask gatekeeper and an energy regulator in skeletal muscle. Acute activation of AMPK has been shown to promote glucose transport and fatty acid oxidation while suppressing glycogen synthase activity and protein synthesis. Chronic activation of AMPK induces a shift in muscle fiber type composition, reduces markers of muscle degeneration and enhances muscle oxidative capacity potentially by stimulating mitochondrial biogenesis. Furthermore, recent evidence demonstrates that AMPK may not only regulate metabolism during exercise but also in the recovery phase. AMPK acts as a molecular transducer between exercise and insulin signaling and is necessary for the ability of prior contraction/exercise to increase muscle insulin sensitivity. Based on these observations, drugs that activate AMPK might be expected to be useful in the treatment of metabolic disorders and insulin resistance in various conditions.

**Introduction**

One fundamental parameter that living cells need to sustain essential cellular functions is the maintenance of sufficiently high level of ATP. Thus, cell survival is dependent on a dynamic control of energy metabolism when ATP demand needs to remain in balance with ATP supply. If ATP consumption exceeds ATP production, the ADP:ATP ratio rises, but this is converted into an even larger rise in AMP:ATP ratio due to the reaction catalyzed by adenylate kinase (2ADP <-> ATP + AMP). If the reaction is at equilibrium, the AMP:ATP ratio will vary as the square of the ADP:ATP ratio, making increases in AMP a more sensitive indicator of energy stress than decreases in ATP or increases in ADP. On this basis, the cell requires an efficient energy sensory mechanism based on the detection of the ratios of ADP:ATP or AMP:ATP. Such a system has been identified as the AMP-activated protein kinase (AMPK), a heterotrimeric serine/threonine kinase conserved throughout eukaryote evolution (Hardie et al. 2012). The primary function of AMPK is to monitor changes in the intracellular level of ATP and maintain energy stores by reprogramming metabolism through an increase in the rate of catabolic ATP-producing pathways and a decrease in the rate of nonessential anabolic ATP-utilizing pathways. These regulatory features are initiated by the phosphorylation of key metabolic enzyme as well as transcription factors for both short-term effects and long-range regulatory actions for a better response to future challenges. Although the AMPK system originally evolved to regulate energy homeostasis in a cell-autonomous manner, in multicellular organisms, its role has adapted to integrate stress responses such as exercise as well as nutrient and hormonal signals to control food intake, energy expenditure, and substrate utilization at the whole body level (Hardie 2014). Activation of AMPK is triggered by a diverse array of external (e.g., exercise, hormones, nutrients) and internal signals (e.g., AMP/ATP and ADP/ATP ratios) and has been implicated in the regulation of a wide range of biochemical pathways and physiological processes. As a consequence, AMPK has stimulated much interest due to its potential impact on metabolic disorders. The aim of this chapter is to discuss the possible role of AMPK in the adaptations to exercise, nutrient and hormonal signals and its potential as a therapeutic drug target, mimicking the beneficial effects of exercise.

**AMPK: structure and regulation**

AMPK is a heterotrimeric complex composed of one catalytic α-subunit comprising a typical Ser/Thr kinase domain, in combination with scaffolding β-subunit containing a carbohydrate binding module (CBM) and γ-subunit containing four cystathionine-β-synthase (CBS) domains that serve to bind adenine nucleotides (Fig. 1). Each of these subunits has several isoforms encoded by different genes (α1, α2, β1, β2, γ1, γ2 and γ3) that can theoretically combine to form 12 possible heterotrimeric complexes (Hardie et al. 2012). Differential expression of the AMPK isoform in tissues and post-translational modifications that may locate AMPK in different cellular compartments also contribute to the specialized functions of AMPK heterotrimeric complexes. Of interest, in contrast to other AMPK isoforms, expression of the AMPKγ3 isoform is restricted to the fast-twitch glycolytic skeletal muscle, suggesting a particular role for AMPKγ3-containing complexes to handle metabolic challenges in skeletal muscle (Barnes et al. 2004). Mutation in the gene encoding for the AMPKγ3 subunit PRKAG3 has been reported in pigs and humans, causing increased deposition of glycogen in skeletal muscle (Milan et al. 2000; Costford et al. 2007). In human skeletal muscle, only three heterotrimeric complexes have been detected, α2β2γ1 (65% of the total pool), α2β2γ3 (20%), and α1β2γ1 (15%) (Birk and Wojtaszewski 2006; Fig. 1). The heterotrimer combination varies in mouse skeletal muscle, with the detection of five complexes, including α1 and α2-associated AMPK complexes with both β1 and β2 isoforms (Treebak et al. 2009). Interestingly, each heterotrimer combination displays a distinct activation profile in response to physical exercise, with γ3-containing complexes being predominantly activated and α2β2γ1 and α1β2γ1 heterotrimers being unchanged or activated only after prolonged exercise (Birk and Wojtaszewski 2006).

The mechanism of AMPK activation involves two steps, a reversible phosphorylation at a conserved residue (Thr174 in α1 and Thr172 in α2 catalytic subunit, hereafter referred to as Thr172) within the activation loop in the α-subunit, and a stimulatory allosteric effect upon binding of AMP within the CBS domains of the γ-subunit (Hardie et al. 2012). Activity of the complex increases more than 100-fold when AMPK is phosphorylated on Thr172 by identified upstream kinases. The combined effect of phosphorylation on Thr172 and allosteric regulation causes a >1,000-fold increase in kinase activity, allowing high sensitivity in responses to small changes in cellular energy status. In addition, AMP and ADP binding regulates AMPK activity by promoting Thr172 phosphorylation by the upstream kinases and by protecting Thr172 from dephosphorylation by phosphatases. All the binding effects of AMP and ADP are antagonized by binding of ATP, providing a very sensitive mechanism for the activation of AMPK in conditions of cellular energy stress. Recent crystallographic studies of full-length AMPK heterotrimeric complexes have provided insights into the domain structure and the regulation upon binding of adenosine nucleotides (Hardie et al. 2016). Because the activating ligand is bound on the γ subunit and the kinase domain is in the α subunit, intersubunit communication has to occur when switching to fully active states. Important regulatory features for this conformational switch are provided by α subunit flexible components (Fig. 1), the autoinhibitory domain (AID) and the α-regulatory subunit interacting motif (α-RIM)/α-hook interacting with the exchangeable nucleotide-binding sites on the γ subunit, offering a signaling mechanism for nucleotide allosteric regulation and protection against dephosphorylation of AMPK heterotrimeric complex.
In mammals, the major upstream kinases are the the liver kinase B1 (LKB1) and $Ca^{2+}$/calmodulin-dependent protein kinase kinase 2 (CaMKK2; Hardie et al. 2012). Interestingly, CaMKK2 has been shown to phosphorylate and activate AMPK in response to an increase in intracellular $Ca^{2+}$ concentration, independent of any change in cellular AMP:ATP or ADP:ATP ratios. In skeletal muscle, the major upstream kinase phosphorylating α subunit Thr172 is liver kinase B1 (LKB1), as exercise-induced AMPK phosphorylation is prevented in mouse models lacking LKB1 (Sakamoto et al. 2005; Thomson et al. 2007). However, CaMKKβ has been shown to activate AMPK during mild tetanic skeletal muscle contraction (Jensen et al. 2007) and to increase AMPKα1 activity in response to skeletal muscle overload in LKB1-deficient mice (McGee et al. 2008).

**AMPK: regulation by hormones and nutrients**

Although AMPK was originally identified as a sensor of cellular energy status by coordinating anabolic and catabolic pathways to balance nutrient supply with energy demand, it is now clear that it also participates in controlling whole-body energy homeostasis by integrating hormonal and nutritional signals from the cellular environment and the whole organism. Hypothalamic AMPK has been suggested to be a key mediator in the regulation of neuronal regulation of feeding behaviour and energy balance (Fig. 2). Inhibition of AMPK by expressing a dominant negative isoform in the arcuate nucleus (ARC) of the hypothalamus decreases mRNA expression of the orexigenic neuropeptides agouti-related peptide (AgRP) and neuropeptide Y (NPY ; Minokoshi et al. 2004). Conversely, activation of AMPK in the ARC by expressing a constitutively active AMPK form can further increase the fasting-induced expression of AgRP and NPY and then elevated feeding (Minokoshi et al. 2004). Similarly, starvation induces AMPK activation and food intake. In contrast, refeeding and glucose administration promote AMPK inactivation, accompanied by increased levels of anorexigenic neuropeptides proopiomelanocortin (POMC) and reduced levels of orexigenic neuropeptides AgRP mRNA in the ARC, highlighting the sensing levels of nutrients in the hypothalamus (Andersson et al. 2004; Minokoshi et al. 2004). There are also a number of hormones involved in the regulation of appetite that alter AMPK activaty in the hypothalamus. For example, the orexigenic hormones, such as ghrelin and adiponectin, activate AMPK in the ARC and promote food intake (Andersson et al. 2004; Kubota et al. 2007); in contrast, anorectic hormones, such as leptin and oestradiol, inhibit AMPK in the ARC and inhibit food intake (Yang et al. 2011; Martinez de Morentin et al. 2014). Another layer of signal integration for the regulation of whole-body energy balance happens at the level of the ventromedial nucleus (VMH) of the hypothalamus, where AMPK regulates energy expenditure by contolling brown adipose tissue (BAT) thermogenesis (Lopez et al. 2010; Whittle et al. 2012; Beiroa et al. 2014; Martinez de Morentin et al. 2014). Importantly, expression of a constitutively active AMPK form in the VMH is associated with a specific reduction in the expression of BAT thermogenic markers. In contrast, inhibition of AMPK by administration of thyroid hormone T3 to the VMH promotes whole body energy expenditure by triggering BAT thermogenesis via activation of the sympathetic nervous system (Lopez et al. 2010). More recently, it was found that injection of glucagon-like peptide-1 (GLP-1) receptor agonist liraglutide into VMH decreased AMPK activity, stimulated expression of thermogenic markers in BAT, and promoted weight loss without affecting food intake (Beiroa et al. 2014). Overall these findings demonstrate a key role for central AMPK in the regulation of energy balance by influencing food intake and energy expenditure in response to peripheral signals, such as hormones and nutrients.

**AMPK: regulation by exercise**

Lifestyle intervention such as regular physical exercise is widely recognized to improve whole-body performance and metabolism in health and disease. An increase in daily physical activity is an effective approach to combat many disease symptoms associated with metabolic syndrome. Endurance exercise can improve insulin sensitivity and metabolic homeostasis. However, our understanding of how exercise exerts these beneficial effects is incomplete. In response to exercise, ATP turnover is increased by more than 100-fold (Gaitanos et al. 1993), resulting in increased ATP consumption and a rise in intracellular AMP levels due to the adenylate kinase reaction. These changes in the adenylate energy charge lead to the activation of AMPK in an intensity- and time-dependent manner, as shown in rodents (Winder and Hardie 1996) and in human muscle (Wojtaszewski et al. 2000). Once activated, AMPK regulates multiple signaling pathways whose overall effects are to increase ATP production, including fatty acid oxidation and glucose uptake. Given that AMPK is at the nexus of metabolic signaling pathways, a great deal of interest has focused on the role of AMPK in the adaptation of skeletal muscle to exercise as well as its use as a possible therapeutic target for the treatment of type 2 diabetes.

Control of exercise-induced glucose transport
During exercise, contracting skeletal muscle rapidly increases glucose uptake in an intensity-dependent manner to sustain the energy demand caused by increased ATP turnover. The first compelling evidence for a role for AMPK in regulating glucose uptake in skeletal muscle has been obtained with pharmacological activation of AMPK by AICAR (Merrill et al. 1997). This finding was further supported by the observation of impaired response to AICAR stimulation in mice expressing a dominant negative AMPK form in skeletal muscle (Mu et al. 2001). However, the role for AMPK in regulating glucose uptake during muscle contractile activity remains controversial and no solid genetic evidence has been put forward. Exercise-induced muscle glucose uptake is impaired in AMPK β1β2M-KO mice but remains intact in AMPKα mdKO (O'Neill et al. 2011; Lantier et al. 2014). The reason for the difference between studies is not clear and future studies using inducible mouse models to study the role of AMPK in developed adult skeletal muscle are warranted.

It has been suggested that AMPK enhances glucose uptake by increasing the translocation of glucose transporter type 4 (GLUT4) to the plasma membrane (Fig. 3). Recent findings using AMPK-deficient mouse models have shown a convergence to the phosphorylation of the downstream target of TBC1D1, Rab-GTPase activating protein, which is emerging as an essential player in contraction-stimulated GLUT4 translocation (Stockli et al. 2015). In support of this finding, mice expressing TBC1D1 that mutated at predicted AMPK phosphorylation sites showed reduced contraction-stimulated glucose uptake (Vichaiwong et al. 2010). Interestingly, in exercised human skeletal muscle, TBC1D1 phosphorylation was significantly correlated with the activity of the α2β2γ3 heterotrimer, supporting the idea that AMPK is a direct upstream TBC1D1 kinase (Treebak et al., 2014).

Recent studies using compound 991, a cyclic benzimidazole derivative and potent direct AMPK activator, have shown that pharmacological activation of AMPK is sufficient to elicit metabolic effects in muscle appropriate for treating type 2 diabetes (Lai et al. 2014). It is also important to note that AMPK-mediated glucose uptake is not impaired in type 2 diabetes during exercise (Musi et al. 2001); therefore, activation of AMPK represents an attractive target for intervention.

Control of training-induced muscle adaptations
In response to repeated metabolic stress, AMPK orchestrates a coordinated response to enhance mitochondrial biogenesis to match substrate utilization to demand (Fig. 3 ; Zong et al., 2002). This response is mediated through the regulation of peroxisome proliferator-activated receptor γ co-activator-1α (PGC-1α), a transcriptional co-activator that promotes the expression of mitochondrial genes encoded in both nuclear and mitochondrial DNA. AMPK activation causes the stimulation of PGC-1α expression by direct phosphorylation, which drives expression from its own promoter but also involves deacetylation of PGC-1α via the silent mating-type information regulator 2 homolog 1 (SIRT1; McGee and Hargreaves 2010). This finding was further supported by the increase seen in PGC-1α and mitochondrial function in mice expressing a constitutively active AMPK form in skeletal muscle (Garcia-Roves et al. 2008). Moreover, deletion of AMPK greatly reduced exercise-induced SIRT1-dependent activation of PGC-1α signaling in skeletal muscle (Canto et al. 2010). Consistent with these findings, reduced muscle AMPK activity has been associated with decreased mitochondrial content during aging (Reznick et al. 2007) and in skeletal muscle from AMPKβ1β2- and LKB1-deficient mice (O'Neill et al. 2011; Tanner et al. 2013).

Control of muscle insulin sensitivity
Skeletal muscle demonstrates increased insulin-stimulated glucose uptake in the period after exercise (Richter et al. 1982). It may involve an increased translocation of GLUT4 at the plasma membrane in response to insulin. The underlying mechanism appears to involve AMPK-dependent phosphorylation of the Rab-GTPase TBC1D4 regulating GLUT4 translocation (Kjobsted et al. 2016). These results are fully in line with previous findings showing that prior pharmacological AMPK activation by AICAR enhances insulin sensitivity in rat skeletal muscle with increased phosphorylation of TBC1D4 (Kjobsted et al. 2015). Thus, activation of AMPK in the recovery after exercise may be important for exercise-induced adaptations and may serve to enhance muscle insulin sensitivity (Fig. 3). These findings may be highly relevant for pharmacological interventions in the treatment of muscle insulin resistance.

**Conclusions and therapeutic perspectives**

In mammals, AMPK has emerged as a major energy sensor that integrates multiple extracellular and intracellular input signals to coordinate cellular energy balance. Targeted by nutritional and hormone signals, AMPK has a crucial role in the hypothalamus to regulate energy intake and energy expenditure. In skeletal muscle, AMPK has been identified as an important integrator of the metabolic changes that occur during physical exercise. Given these roles, AMPK is an obvious target for treatment of metabolic disorders such as obesity and diabetes. However, although pharmacological activation in peripheral organs is expected to provide therapeutic benefits, activating central AMPK could cause deleterious consequences, specifically on body weight control. Recent studies have highlighted the adverse metabolic consequence of AMPK activation throughout all tissues (Yavari et al. 2016). Thus, a better understanding of the precise role of AMPK in the cell and its effects at the whole-body level will be essential for delineating therapeutic strategies aimed at targetting AMPK.

**Figure legends:**

**Fig. 1.** Schematic representation of AMPK subunit. (**A**) AMPK domain structure. (**B**) AMPK heterotrimer composition in human skeletal muscle.

**Fig. 2.** Hypothalamic AMPK in the regulation of energy balance.

**Fig. 3.** AMPK-mediated regulation of skeletal muscle adaptation to exercise